\documentclass[prd,aps,preprint,11pt,onecolumn]{revtex4-1}

\usepackage{amsfonts,amsmath,amssymb,graphicx,bm,subfigure}
\usepackage[colorlinks,
  linkbordercolor={0 0 0},
  pdfborder={0 0 0},
  linkcolor=blue,
  citecolor=blue,
  urlcolor=blue,
  breaklinks=true]{hyperref}

\begin{document}

\author{Mridupawan Deka$^{1)}$}
\email{mpdeka@theor.jinr.ru}

\author{Maxim Dvornikov$^{1), 2)}$}
\email{maxim.dvornikov@gmail.com}

\title{Neutrino spin oscillations near a black hole}

\affiliation{$^{1)}$\,Bogoliubov Laboratory of Theoretical Physics, Joint Institute for Nuclear Research, Dubna, Russia \\
$^{2)}$\,Pushkov Institute of Terrestrial Magnetism, Ionosphere and Radiowave Propagation (IZMIRAN), Troitsk, Moscow, Russia}

\begin{abstract}
  In this work, we study neutrino spin oscillations in the case when they are
  gravitationally scattered off a rotating Kerr black hole surrounded by a
  thick magnetized accretion disk. We consider only toroidal magnetic field
  inside the disk. Neutrino spin precession is caused by the interaction of
  the neutrino magnetic moment with the magnetic field in the disk.
  Our treatment of the spin oscillations of the observed neutrino fluxes
  is based on numerical simulations of the propagation of a large number of
  incoming test neutrinos using High Performance Parallel
  Computing. We briefly discuss our results and their applications in the
  observations of astrophysical neutrinos.
\end{abstract}

\maketitle

\section{Introduction}
\label{sec:INTR}

Neutrinos are experimentally confirmed to have non-zero masses, and there is a
mixing between different types of neutrino resulting in flavor oscillations (see,
e.g., refs.~\cite{NOvA:2021nfi,GiuKim07}). In addition, neutrinos are believed
to have non-zero magnetic moment~\cite{Beda:2012zz, Dvornikov:2003js,
  Broggini:2012df, Giu16}. Such a non-zero magnetic moment of neutrino leads to
neutrino spin oscillations due to the electromagnetic and electroweak interactions
of neutrinos with the external fields and background
matter~\cite{Fujikawa:1980yx}. Spin oscillations result in left polarized
active neutrinos being converted to sterile right handed ones of the same flavor. 

In addition to photons, it is found that a significant flux of neutrinos are
emitted by an accretion disk of a black hole (BH)~\cite{Caballero:2011dw}.
The observation of gravitationally scattered neutrinos off a BH surrounded by
an accretion disk gives one a possibility to examine the neutrino
spin oscillations. These neutrinos experience electromagnetic interactions with
the magnetic field(s) of the accretion disk as well as electroweak interactions
with the background matter of the disk~\cite{Okun:1986na}. As a result, their
spins precess leading to spin oscillations.

Using the quasi-classical approach for the study of the motion of the spinning
particles described in Ref.~\cite{PomKhr98}, neutrino spin oscillations in
curved spacetime within General Relativity are studied in Ref.~\cite{Dvo06}.
The contributions of the neutrino electromagnetic and electroweak interactions
have accordingly been incorporated inside this approach in Ref.~\cite{Dvo13}.
Extensive studies of the spin oscillations of the gravitationally scattered
neutrinos have then been performed in
Refs.~\cite{Dvo23c,Dvo23d,Dvo23a,Dvo23b,Deka:2023ljj}. In a
gravitational scattering, both ``in'' and ``out'' states of neutrinos are in
the asymptotically flat spacetime. Therefore, their spin states are well
defined.

In the present work, following Refs.~\cite{Dvo23b,Deka:2023ljj}, we consider a
rotating supermassive BH (SMBH) surrounded by a thick magnetized accretion disk.
The model of such a disk has been proposed in Ref.~\cite{Abramowicz_1978} called
a ``Polish doughnut''.  The toroidal magnetic field has been introduced in the
disk in Ref.~\cite{Kom06}.

The main difference in this work is that we consider
only toroidal magnetic field in the disk unlike in
Refs.~\cite{Dvo23b,Deka:2023ljj} where both toroidal and poloidal magnetic
fields have been considered together. In comparison to Ref.~\cite{Dvo23b}, we
use a few million incoming test neutrinos, with the help of High Performance
Parallel Computing, to significantly increase the resolution of fluxes of the
scattered neutrinos.

This work is organized in the following way. First, in Sec.~\ref{sec:formalism},
we briefly outline our approach. In Secs.~\ref{subsec:MOTION}
and~\ref{subsec:SPIN_EV}, we discuss the motion and the spin evolution of a test
particle in the gravitational field of a rotating BH, respectively. Also, the
interactions of the neutrino spin with the toroidal magnetic fields and background
matter are discussed. The numerical methods and parameters are described in
Sec.~\ref{sec:NUMERICAL}. The results are then presented in Sec.~\ref{sec:RES}.
Finally, we conclude our work in Sec.~\ref{sec:CONCL}.

\section{Formalism}
\label{sec:formalism}

We suppose that a beam of incoming neutrinos, which are emitted from a source with the
coordinate, $(r,\theta,\phi)_s = (\infty,\pi/2,0)$, is traveling in the vicinity
of a spinning SMBH. While some of them fall into the BH, the rest 
are scattered gravitationally. Since we assume that a neutrino has a nonzero magnetic
moment $\mu$, the scattered beam of neutrinos experience electromagnetic
interactions with the toroidal magnetic field of the magnetized accretion disk
surrounding the SMBH as well as electroweak interactions with the matter fields
in the accretion disk. These interactions result in neutrino spin precession. The neutrinos
eventually escape to infinity, and finally are observed at the position,
$(r,\theta,\phi)_{\mathrm{obs}} = (\infty,\theta_\mathrm{obs},\phi_\mathrm{obs})$.
Our goal is to study the probability distributions of spin precession of these
neutrinos as functions of $\theta_\mathrm{obs}$ and $\phi_\mathrm{obs}$. Below we
define the trajectories and spin evolution of these neutrinos between
$(r,\theta,\phi)_s$ and $(r,\theta,\phi)_{\mathrm{obs}}$. 

\subsection{Motion of ultra-relativistic neutrinos in Kerr Spacetime}
\label{subsec:MOTION}

The spacetime of a spinning BH is described in Kerr metric. For a BH with the
mass $M$ and angular momentum $J$ along the $z$-axis, the metric can be written
in Boyer\---Lindquist coordinates, $x^{\mu}=(t,r,\theta,\phi)$, as,
\begin{equation}\label{eq:Kerrmetr}
  \mathrm{d}s^{2}=g_{\mu\nu}\mathrm{d}x^{\mu}\mathrm{d}x^{\nu}=
  \left(
    1-\frac{rr_{g}}{\Sigma}
  \right)
  \mathrm{d}t^{2}+2\frac{rr_{g}a\sin^{2}\theta}{\Sigma}\mathrm{d}t\mathrm{d}\phi-\frac{\Sigma}{\Delta}\mathrm{d}r^{2}-
  \Sigma\mathrm{d}\theta^{2}-\frac{\Xi}{\Sigma}\sin^{2}\theta\mathrm{d}\phi^{2},
  \notag
\end{equation}
where,
\begin{equation}\label{eq:dsxi}
  \Delta=r^{2}-rr_{g}+a^{2},
  \quad
  \Sigma=r^{2}+a^{2}\cos^{2}\theta,
  \quad
  \Xi=
  \left(
    r^{2}+a^{2}
  \right)
  \Sigma+rr_{g}a^{2}\sin^{2}\theta. \notag
\end{equation}
Here, $r_g = 2M$ is the Schwarzschild radius and $J = a M$ where $0 < a < M$ .

The geodesic motion of an ultra-relativistic test particle in Kerr metric has three
constants of motion: the particle energy, $E$, its angular momentum, $L$, and the
Carter constant, $Q$. Since we are considering scattering only, therefore $Q>0$.
The trajectory of an ultra-relativistic neutrino in the presence of the
gravitational field of a spinning SMBH can be written
as~\cite{GraLupStr18},
\begin{align}
  & \int\frac{\mathrm{d}r}{\pm\sqrt{R}}=\int\frac{\mathrm{d}\theta}{\pm\sqrt{\Theta}},
  \label{eq:trajth}
  \\
  & \phi = a\int\frac{\mathrm{d}r}{\pm\Delta\sqrt{R}}[(r^{2}+a^{2})E-aL]+
  \int\frac{\mathrm{d}\theta}{\pm\sqrt{\Theta}}
  \left[
    \frac{L}{\sin^{2}\theta}-aE
  \right].
  \label{eq:trajphi}
\end{align}
where $R$ and $\Theta$ potentials are defined as,
\begin{align}
  R(r) = & [(r^{2}+a^{2})E-aL]^{2}-\Delta[Q+(L-aE)^{2}],\notag
  \\
  \Theta(\theta)= & Q + \cos^{2}\theta
  \left(
    a^{2} E^{2} - \frac{L^{2}}{\sin^{2}\theta} \notag
  \right). \notag
\end{align}
We choose $\pm{}$ signs for $\sqrt{R}$ and $\sqrt{\Theta}$ in
Eqs.~\eqref{eq:trajth} and~\eqref{eq:trajphi} to be same as those of
$\mathrm{d}r$ and $\mathrm{d}\theta$ depending upon whether a neutrino
approaches or moves away from the BH.

By defining the dimensionless variables, $r=xr_{g}$, $L=yr_{g}E$, $Q=wr_{g}^{2}E^{2}$,
and $a=zr_{g}$, we can write the radial integral from any point $x$ to $\infty$
as,
\begin{align}\label{eq:Ix}
  I_x = & z \sqrt{t_+^2 + t_-^2}
  \int_x^\infty \frac{\mathrm{d}x'}{\sqrt{x'^{4}+x'^{2}\left[z^{2}-w-y^{2}\right]+x'\left[w+\left(z-y\right)^{2}\right]-z^{2}w}},
\end{align}
where,
\begin{align}
  t_{\pm}^{2}=\frac{1}{2z^{2}}\left[\sqrt{(z^{2}-y^{2}-w)^{2}+4z^{2}w}\pm(z^{2}-y^{2}-w)\right].\notag
\end{align}

For an incoming neutrino moving above the equatorial plane of the BH from infinity,
the number of inversions of its trajectory with respect to the equatorial plane
is given by
\begin{equation}\label{eq:Ninv_in}
  N =
  \left\lfloor
    \frac{1}{2}
    \left(
      \frac{I_x}{K} -1
    \right)  
   \right\rfloor
   +1,
\end{equation}
where, $\lfloor\cdots\rfloor$ denotes the Floor function, and
$K = K \left(\frac{t_{+}^{2}}{t_{-}^{2}+t_{+}^{2}}\right)$ is the
complete elliptic integral of the first kind. We follow the definitions of
elliptic integrals and functions corresponding to those in Ref.~\cite{AbrSte64}.

Using Eqs.~\eqref{eq:trajth}, \eqref{eq:trajphi}, \eqref{eq:Ix}
and~\eqref{eq:Ninv_in}, we can find $\theta(x)$ and $\phi(x)$ for the
incoming neutrinos from the following relations,
\begin{eqnarray}
  \cos\theta &=& t_+ \text{cn}
  \left(
  (-1)^N \left\{ K \left( \frac{t_{+}^{2}}{t_{-}^{2}+t_{+}^{2}} \right)
  \left(4 \left\lfloor \frac{N}{2} \right\rfloor +1 \right) - I_x \right\} \bigg| \frac{t_{+}^{2}}{t_{-}^{2}+t_{+}^{2}}
  \right),
  \label{eq:thetabtp}
   \\
   \phi_{\mathrm{in}} &=& z \displaystyle\int^\infty_x
  \frac{(x-zy) dx} {(x^2 -x + z^2) \sqrt{R(x)}}
  + \frac{y}{zt_-}
  \left\{2 N \Pi + (-1)^N \Pi_t\right\},
  \label{eq:phibtp}
\end{eqnarray}
where $\text{cn}(n|m)$ is the elliptic Jacobi function, and $t = \cos\theta(x)$
at $x$. $\Pi$ and $\Pi_t$, respectively, are the complete and incomplete elliptic
integrals of third kind,
\begin{equation}
  \Pi = \Pi\left(t_{+}^{2}\left|\right. - t_{+}^{2}/t_{-}^{2}\right),\hspace{2mm}
  \Pi_t = \Pi\left(t_{+}^{2}, \arcsin(t/t_+)\left|\right. - t_{+}^{2}/t_{-}^{2}\right).
  \notag
\end{equation}

When the neutrino moves from the turn point towards the infinity, then the number
of inversions is given by
\begin{equation}\label{eq:Ninv_out}
\displaystyle N =
  \begin{cases}
  \left\lfloor
    \frac{I_x + F}{2K} 
   \right\rfloor, & \text{if} \quad \dot{t}_\mathrm{tp} < 0,
  \\
  \left\lfloor
    \frac{I_x - F}{2K} 
   \right\rfloor + 1, & \text{if} \quad \dot{t}_\mathrm{tp} >0,
  \end{cases}
\end{equation}
where ${t}_\mathrm{tp} = \cos \theta(\mathrm{tp})$ at the turn point, and $F$ is
the incomplete elliptic integral of first kind,
\begin{equation}%\label{eq:Ninv}
  F = F
  \left(
    \arccos \left( \frac{t_\mathrm{tp}}{t_{+}} \right),\frac{t_{+}^{2}}{t_{-}^{2}+t_{+}^{2}}
  \right).\notag
\end{equation}

Using Eqs.~\eqref{eq:trajth}, \eqref{eq:Ix} and~\eqref{eq:Ninv_out}, we
can find $\theta(x)$ for the outgoing neutrinos as,
\begin{equation}
  \cos\theta =
  t_+\times
  \begin{cases}
    \text{cn}
    \left(
      (-1)^N
      \left(
        I_x + F - 4 K
        \left\lceil
          \frac{N}{2}
        \right\rceil
      \right)
      \Big| \frac{t_{+}^{2}}{t_{-}^{2}+t_{+}^{2}}
    \right),
    & \text{if} \quad \dot{t}_\mathrm{tp} < 0,
    \\
    \text{cn}
    \left(
      (-1)^N
      \left(
        F - I_x + 4 K
        \left\lfloor
          \frac{N}{2}
        \right\rfloor
      \right)
      \Big| \frac{t_{+}^{2}}{t_{-}^{2}+t_{+}^{2}}
    \right),
  & \text{if} \quad \dot{t}_\mathrm{tp} >0.
  \end{cases}
  \label{eq:thetaatp}
\end{equation}
Similarly using Eqs.~\eqref{eq:trajphi}, \eqref{eq:Ix} and~\eqref{eq:Ninv_out},
we find $\phi(x)$ for outgoing neutrions as,
\begin{equation}
  \phi_{\mathrm{out}} =
  \begin{cases}
    z \displaystyle\int^\infty_x
    \frac{(x-zy) dx} {(x^2 -x + z^2) \sqrt{R(x)}}
    + \frac{y}{zt_-}
    \left\{2 N \Pi + \Pi_{\mathrm{tp}} - (-1)^N \Pi_t\right\},
    & \text{if} \quad \dot{t}_\mathrm{tp} < 0,
    \\
    z \displaystyle\int^\infty_x
    \frac{(x-zy) dx} {(x^2 -x + z^2) \sqrt{R(x)}}
    + \frac{y}{zt_-}
    \left\{2 N \Pi - \Pi_{\mathrm{tp}} + (-1)^N \Pi_t\right\},
    & \text{if} \quad \dot{t}_\mathrm{tp} > 0.
    \label{eq:phiatp}
  \end{cases}
\end{equation}
Here $\Pi_{\mathrm{tp}}= \Pi\left(t_{+}^{2},
\arcsin(t_{\mathrm{tp}}/t_+)\left|\right. - t_{+}^{2}/t_{-}^{2}\right)$.

In a similar manner, we can also describe the motion of incoming neutrinos which
propagate below the equatorial plane of the BH.

Note that $\phi_\mathrm{obs} = \phi_{\mathrm{in}} + \phi_{\mathrm{out}}$. Also,
a neutrino can make multiple revolutions around the BH. This results in the
azimuthal angle, $\phi$, being greater than $2\pi$. One should account for
it in the final analysis.

\subsection{Neutrino polarization in the presence of Accretion Disk}
\label{subsec:SPIN_EV}

As mentioned earlier, a neutrino interacts electroweakly with the accretion disk
besides gravity. In addition, it experiences spin precession due to
the electromagnetic interaction with the magnetized disk~\cite{Kom06}. In this
section, we shall discuss the evolution of neutrino polarization along the
trajectories described in Sec.~\ref{subsec:MOTION}.

The polarization of neutrino can be described by an invariant three vector
$\bm{\zeta}$ in the rest frame with respect to a locally Minkowskian frame.
The evolution of the neutrino polarization vector obeys,
\begin{equation}\label{eq:nuspinrot}
  \frac{\mathrm{d}\bm{\bm{\zeta}}}{\mathrm{d}t}=2(\bm{\bm{\zeta}}\times\bm{\bm{\Omega}}),
\end{equation}
where,
\begin{equation}
  {\bm{\Omega}}={\bm{\Omega}}_{\mathrm{g}}+{\bm{\Omega}}_{\mathrm{em}}+{\bm{\Omega}}_{\mathrm{matt}}. \notag
\end{equation}
The explicit forms of the gravitational, electromagnetic and electroweak
interactions, ${\bm{\Omega}}_{\mathrm{g,em,matt}}$, respectively, are given in
Refs.~\cite{Dvo23a,Dvo23b}.

However instead of dealing with Eq.~\eqref{eq:nuspinrot}, it is more numerically
convenient to study the effective Schr\"odinger equation for the description of the
neutrino polarization,
\begin{equation}\label{eq:Schreq}
  \mathrm{i}\frac{\mathrm{d}\psi}{\mathrm{d}x}= \hat{H}_{x}\psi,
\end{equation}
where,
\begin{equation}
  \hat{H}_{x}= -\mathcal{U}_{2}(\bm{\bm{\sigma}}\cdot\bm{\bm{\Omega}}_{x})\mathcal{U}_{2}^{\dagger},
  \quad
  \bm{\bm{\Omega}}_{x} =  r_{g}\bm{\bm{\Omega}}\frac{\mathrm{d}t}{\mathrm{d}r},
  \quad
  \mathcal{U}_{2}=\exp(\mathrm{i}\pi\sigma_{2}/4).\notag
\end{equation}
Here $\bm{\bm{\sigma}}=(\sigma_{1},\sigma_{2},\sigma_{3})$ are the Pauli matrices.
The Hamiltonian $\hat{H}_{x}$ is the function of $x$ only through the dependence of
$\theta(x)$ given in Eqs.~\eqref{eq:thetabtp} and~\eqref{eq:thetaatp}. The initial
condition has the form, $\psi_{-\infty}^{\mathrm{T}}=(1,0)$, which means all incoming
neutrinos are left polarized. The solution of Eq.~\eqref{eq:Schreq} provides the
polarization of a scattered neutrino in the form,
$\psi_{+\infty}^{\mathrm{T}}=(\psi_{+\infty}^{(\mathrm{R})},\psi_{+\infty}^{(\mathrm{L})})$.
The probability that a neutrino remains left polarized at the observer position,
is $P_{\mathrm{LL}}=|\psi_{+\infty}^{(\mathrm{L})}|^{2}$.

\section{Numerical Methods and Parameters}
\label{sec:NUMERICAL}

In order to determine $P_{\mathrm{LL}}$, we solve the Eq.~\eqref{eq:Schreq}
numerically at each spatial point $x$. We solve separately for the neutrinos
traveling either above or below the equatorial plane of the BH. We
first use $4$th order Adam\---Bashforth predictor method for an irregular grid to
obtain the approximate solutions at each $x$. We then use Adam\---Moulton corrector
method to iteratively improve upon the solutions with an appropriate convergence
condition. The only initial condition that is needed is that all incoming
neutrinos, above and below the equatorial plane, are left-handed at infinity
as mentioned earlier in Sec.~\ref{subsec:SPIN_EV}.

After finding
$\psi_{+\infty}^{\mathrm{T}}=(\psi_{+\infty}^{(\mathrm{R})},\psi_{+\infty}^{(\mathrm{L})})$
numerically, we then compute $P_{\mathrm{LL}}$ at the observer position. The angular
co-ordinates $\theta_{\mathrm{obs}}$ can be computed from Eq.~\eqref{eq:thetaatp}
and $\phi_{\mathrm{obs}}$ from Eqs.~\eqref{eq:phibtp} and~\eqref{eq:phiatp}.

In our study, we fix the mass of SMBH at $M = 10^8 M_\odot$. We consider two
different spins of SMBH, namely $a\,=\,2\times 10^{-2} M$ and $0.98 M$. We
assume a thick accretion disk called a ``Polish doughnut''~\cite{Abramowicz_1978}.
The accretion disk consists of a hydrogen plasma, and it rotates around the SMBH
with a relativistic velocity~\cite{Kom06}. Given the mass of SMBH, the  maximal
number density of electrons is taken to be
$n_e^{(\mathrm{max})} = 10^{18}\,\text{cm}^{-3}$~\cite{Jia19}. The disk inherently
contains a toroidal magnetic field~\cite{Kom06}. We set the maximal strength of
the toroidal field at $320\,\text{G}$, which is $\sim 1\%$ of the Eddington limit
for this type of SMBH~\cite{Bes10}.

Here we would like to emphasize that unlike in the  analogous
studies~\cite{Dvo23c,Dvo23d,Dvo23a,Dvo23b,Deka:2023ljj} where \texttt{MATLAB} based
codes were used, the numerical studies of this work are based on a freshly written
\texttt{C++} code which is more efficient, optimized and flexible. Although it
closely follows the old \texttt{MATLAB} code, it is a totally independent code.
In the process of rewriting the code in \texttt{C++}, we have realized a
significant incorrectness in the results produced by the old \texttt{MATLAB}
code when we considered only toroidal field in the accretion disk. The values of
two dimensionless coefficients,
$V_m = \displaystyle\frac{G_F}{\sqrt{2}m_p r^3_g} \sim O(10^{-87})$ and
$V_B = \displaystyle\left(\frac{\mu}{r_g}\right)^2 \sim O(10^{-76})$
(e.g. see Ref~\cite{Dvo23b}), turned out to be too low
for \texttt{MATLAB}'s default precision. As a result, the \texttt{MATLAB}
code produced the numerical values of plasma density and magnetic pressure to be
zeros. This issue is ratified in this
work, and we present new results by considering only toroidal field in the
accretion disk.

\begin{table}
  \caption{Number of neutrinos for different BH spins.}
  \begin{center}\setlength{\tabcolsep}{8pt}
    \begin{tabular}{|c|c|c|}
      \hline\hline
      & {\boldmath$a = 0.02M$}
      & {\boldmath$a = 0.98M$}\\
      &({Fig.~\ref{fig:f1a_toroid}})
      &({Fig.~\ref{fig:f1b_toroid}})\\
      \hline
      {\bf Above the equatorial plane}
      & $1\, 300\, 000$
      & $1\, 650\, 000$\\
      \hline
      {\bf Below the equatorial plane}
      & $1\, 300\, 000$
      & $1\, 650\, 000$\\
      \hline
      {\bf Total}
      & $2\, 600\, 000$
      & $3\, 300\, 000$\\
      \hline\hline
    \end{tabular}
  \end{center}
  \label{tab:number_of_neutrino}
\end{table}

The magnetic moment of the Dirac neutrinos is taken to be
$\mu = 10^{-13}\,\mu_\mathrm{B}$ where $\mu_\mathrm{B}$ is the Bohr magneton. This
value is less than the
astrophysical upper bound of the magnetic moment~\cite{Via13}. We assume that the
neutrinos undergo electroweak interaction with the plasma of the accretion disk
in the forward scattering approximation~\cite{DvoStu02}. We also assume the neutrino spin
oscillations within one neutrino generation, i.e. we suppose that only the diagonal magnetic moment is present.

For our numerical work, we use more than $288$ cores of SkyLake processors
in Govorun super-cluster. The number of neutrinos we use in each case of BH spin
is more than $2$ million. The detailed numbers are listed in
Table~\ref{tab:number_of_neutrino}.

\begin{figure}[htbp]
\centering
\subfigure[]
 {\label{fig:f1a_toroid}
   \includegraphics[width=0.47\hsize]{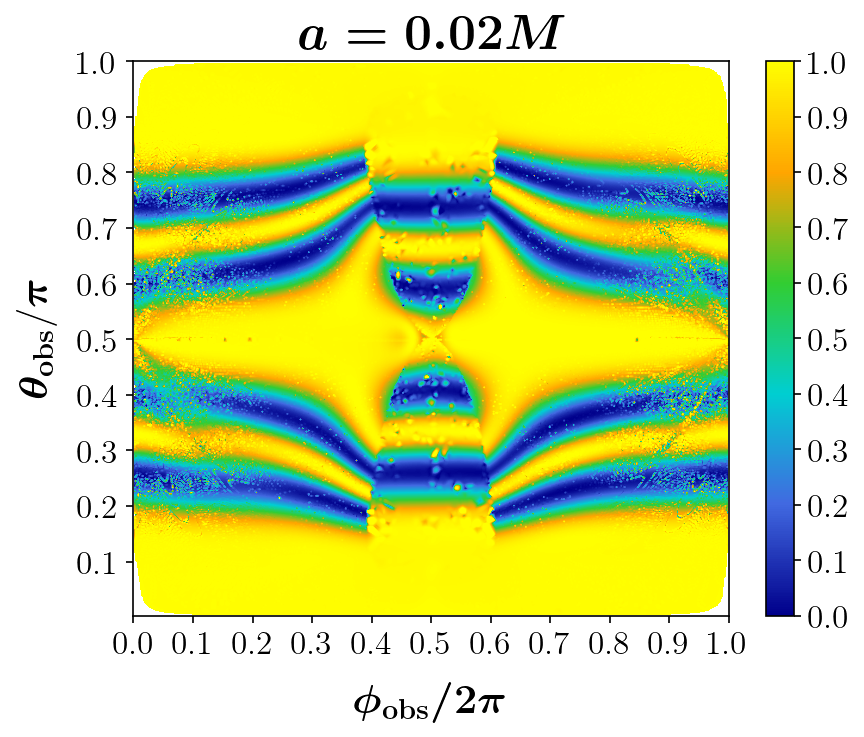}
 }
 \hspace{0mm}
 \subfigure[]
  {\label{fig:f1b_toroid}
    \includegraphics[width=0.47\hsize]{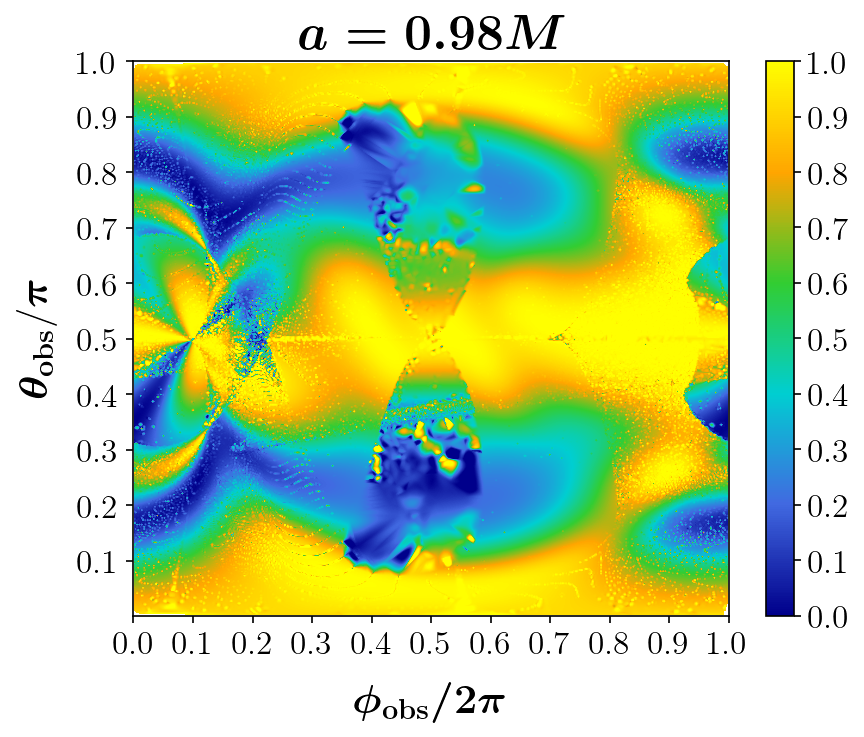}
  }          
  \caption{Contour projections of $P_{\mathrm{LL}}$ as
    functions of $\theta_{\mathrm{obs}}$ and $\phi_{\mathrm{obs}}$.
    The BH spins are,
    (a) $a = 2\times10^{-2}M$
    and
    (b) $a = 0.98M$. BH mass is, $M=10^{8}M_{\odot}$, in both cases.}
\end{figure}

\section{Results}
\label{sec:RES}

If the spin of a left-handed Dirac neutrino precesses in an external field, it
becomes sterile or right-handed. Such a neutrino cannot be observed in a detector.
Hence, the observed flux of neutrinos will be reduced by a factor of
$P_{\mathrm{LL}}$ in comparison to the flux of non-spinning particles.

We present our results in Figs.~\ref{fig:f1a_toroid} and \ref{fig:f1b_toroid}
as functions of $\theta_{\mathrm{obs}}$ and $\phi_{\mathrm{obs}}$ corresponding to
$a\,=\,2\times 10^{-2} M$ and $0.98 M$, respectively. The incoming neutrinos are
traveling in parallel either above or below the equatorial plane of the BH. As
mentioned earlier, we consider only toroidal magnetic field in the accretion disk. All
the areas with $P_{\mathrm{LL}} < 1$ signify spin-flip of neutrinos. The lower the
value of $P_{\mathrm{LL}}$, the higher is the probability of the spin-flip. We see
that in both the spin cases, $a\,=\,2\times 10^{-2} M$ and $0.98 M$, there are
a non-negligible probability of spin-flip.

The result we obtain is different from what was reported in the
studies~\cite{Dvo23c,Dvo23d,Dvo23a,Dvo23b,Deka:2023ljj}. In particular, we get that a sizable neutrino spin-flip takes place even in presence of toroidal magnetic field contrary to Refs.~\cite{Dvo23c,Dvo23d,Dvo23a,Dvo23b,Deka:2023ljj} where a poloidal component is considered.
As mentioned in
Sec.~\ref{sec:NUMERICAL}, this is due to the precision issue involved with
the \texttt{MATLAB} code that was used in those studies. The default numerical
precision of \texttt{MATLAB} was unable to deal with the smallness of the
two dimensionless coefficients, $V_m$ and $V_B$. We have corrected this
issue in our new \texttt{C++} code.

\begin{figure}[htbp]
  {\includegraphics[width=0.48\hsize]{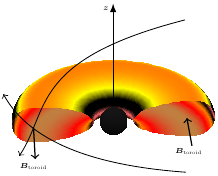}}
  \caption{Schematic diagram showing neutrino trajectories
  which cross the accretion disk. The toroidal magnetic field is perpendicular
  to their momenta.
  \label{fig:tourus_4}}
\end{figure}

A neutrino can undergo spin oscillations when it interacts with a magnetic field
transverse to its velocity. One can see in Fig.~\ref{fig:tourus_4} that there exist
neutrinos which cross the accretion disk while the toroidal magnetic field is
perpendicular to their momenta.

\section{Conclusion}
\label{sec:CONCL}

In this work, we consider the propagation of the ultra-relativistic neutrinos with
nonzero magnetic moment in the strong gravitational field of SMBH. The SMBH is  
surrounded by a thick magnetized accretion disk. These neutrinos are  scattered
due to strong gravitational field of the SMBH. We can exactly describe the
geodesic motion of these neutrinos in Kerr metric.

The neutrinos interact electroweakly with the rotating matter of the accretion
disk. Due to the nonzero magnetic moment, they also interact magnetically with
the disk. In this study, we consider only toroidal field inside the disk. The
neutrino interaction with the external fields causes the spin precession,
which can be accounted for along each neutrino trajectory.

To have a good resolution of $P_{\mathrm{LL}}$ in
the $(\theta_{\mathrm{obs}}, \phi_{\mathrm{obs}})$ plane, we need a large number of
neutrinos. We consider more than $2$ million neutrinos in each case of BH spin.

From the Figs.~\ref{fig:f1a_toroid} and \ref{fig:f1b_toroid}, we clearly see that
there are non-zero probabilities that neutrino spin-flip happens in the presence
of toroidal magnetic field. This is irrespective of the BH spin. This result is in
contrast to Refs.~\cite{Dvo23c,Dvo23d,Dvo23a,Dvo23b,Deka:2023ljj}. The reasons
for the differences in results are explained in Sec.~\ref{sec:NUMERICAL}.

The obtained results can be used in the neutrino tomography of magnetic fields in
the vicinity of BHs. Scattered neutrinos, emitted, e.g., in a supernova explosion,
can be observed by the existing or future neutrino telescopes (see, e.g.,
Ref.~\cite{Abu22}).

\begin{acknowledgments}
  All our numerical computations have been performed at Govorun super-cluster at
  Joint Institute for Nuclear Research, Dubna.
\end{acknowledgments}

\end{document}